\documentclass[twocolumn]{aastex63}
\usepackage{float,graphicx,amsmath,multirow}
\usepackage{color}
\usepackage[version=4]{mhchem}
\usepackage{booktabs}
\usepackage{gensymb}

\graphicspath{{figures/}}

\begin{document}

\title{Machine Learning of Interstellar Chemical Inventories}
\author{Kin Long Kelvin Lee}
\affiliation{Department of Chemistry, Massachusetts Institute of Technology, Cambridge, MA 02139, USA}
\affiliation{Center for Astrophysics $\mid$ Harvard~\&~Smithsonian, Cambridge, MA 02138, USA}
\author{Jacqueline Patterson}
\affiliation{Indiana University, Bloomington, IN 47405, USA}
\affiliation{Center for Astrophysics $\mid$ Harvard~\&~Smithsonian, Cambridge, MA 02138, USA}
\author{Andrew M. Burkhardt}
\affiliation{Center for Astrophysics $\mid$ Harvard~\&~Smithsonian, Cambridge, MA 02138, USA}
\author{Vivek Vankayalapati}
\affiliation{The University of Utah, Salt Lake City, UT 84112, USA}
\affiliation{Center for Astrophysics $\mid$ Harvard~\&~Smithsonian, Cambridge, MA 02138, USA}
\author{Michael C. McCarthy}
\affiliation{Center for Astrophysics $\mid$ Harvard~\&~Smithsonian, Cambridge, MA 02138, USA}
\author{Brett A. McGuire}
\affiliation{Department of Chemistry, Massachusetts Institute of Technology, Cambridge, MA 02139, USA}
\affiliation{National Radio Astronomy Observatory, Charlottesville, VA 22903, USA}
\affiliation{Center for Astrophysics $\mid$ Harvard~\&~Smithsonian, Cambridge, MA 02138, USA}

\correspondingauthor{Kin Long Kelvin Lee, Brett A. McGuire}
\email{kelvlee@mit.edu, brettmc@mit.edu}

\begin{abstract}

The characterization of interstellar chemical inventories provides valuable insight into the chemical and physical processes in astrophysical sources. The discovery of new interstellar molecules becomes increasingly difficult as the number of viable species grows combinatorially, even when considering only the most thermodynamically stable. In this work, we present a novel approach for understanding and modeling interstellar chemical inventories by combining methodologies from cheminformatics and machine learning. Using multidimensional vector representations of molecules obtained through unsupervised machine learning, we show that identification of candidates for astrochemical study can be achieved through quantitative measures of chemical similarity in this vector space, highlighting molecules that are most similar to those already known in the interstellar medium. Furthermore, we show that simple, supervised learning regressors are capable of reproducing the abundances of entire chemical inventories, and predict the abundance of not yet seen molecules. As a proof-of-concept, we have developed and applied this discovery pipeline to the chemical inventory of a well-known dark molecular cloud, the Taurus Molecular Cloud 1 (TMC-1); one of the most chemically rich regions of space known to date. In this paper, we discuss the implications and new insights machine learning explorations of chemical space can provide in astrochemistry.

\end{abstract}

\section{Introduction}
\label{intro}

In the interstellar medium, molecules act as sensitive probes of their local environment. Their relative abundances can be used to infer myriad physical properties of the target system ranging from the thermal history of the source \citep{2002ApJ...571L..55L}, to the kinematic structure of the gas \citep{2018ApJ...860L..13P,2020arXiv200904345D}, the passage of recent hydrodynamic shock \citep{Schilke:1997dt}, or the presence of various radiation fields \citep{2017ApJ...843L...3C}.  The chemical inventories---and abundances---are also deeply tied to the environment itself, from carbon- and silicon-rich evolved stars \citep{Gong:2015ks}, to organic-rich star-forming cores \citep{Belloche:2019hc}, to the salty disks around massive stars \citep{Ginsburg:2019fu}. Thus, the utility of molecules as tracers of the chemical and physical properties and evolutionary history of astrophysical sources increases with the completeness of chemical inventories in these regions: the more knowledge we possess of the inventory, the more astrophysical information we can derive.  

Detecting new molecules in space, and using these detections to infer astrophysics, has underpinned the foundation of astrochemical research for over half a century, although the pace of discovery truly exploded with the advent of molecular radio astronomy in the 1960s \citep{McGuire:2018mc}.  Growing alongside laboratory and observational efforts to identify new molecules in space, astrochemical models were developed to attempt to reconstruct the network of chemical reactions occurring in the environments that were being studied (see, e.g., \citealt{wakelam_2014_2015} and \citealt{Garrod:2008tk}).  The analysis and refinement of these models, from relatively simple networks focusing on just a few molecules \citep{herbst_formation_1973,Guzman:2015iv} to very large holistic approaches attempting to replicate the observed abundances in a source \citep{van_dishoeck_comprehensive_1986,Garrod:2013id}, provide a small window into the complex processes occurring toward a diverse set of environments \citep{Schilke:1997dt}.  Yet, often these models are descriptive rather than predictive: while they replicate many of the abundance ratios seen in an observation and provide substantial insight into the associated chemistry and physics, they can also struggle to predict abundances of other species that have not yet been observed (see, e.g., \citealt{McGuire:2015bp}). For each newly detected molecule, we must contemplate a host of related species and reaction pathways necessary to describe its chemistry. To date, this aspect of astrochemistry has solely depended on chemical intuition: we draw on subjective expertise to determine what \emph{may} be important to dedicate computational, laboratory, and observational efforts to. As the complexity of molecules grows beyond a few carbon atoms, however, the number of possible isomers grows combinatorially and inference based on human intuition becomes neither tractable nor exhaustive. 

% Ultimately, expanding our knowledge of chemical inventories in space usually relies upon chemical intuition: a researcher looking at the collection of molecules seen in a source and making an intuitive leap to what other species might be present.  We currently lack a robust, computer-based methodology for suggesting new species of interest for us to study with computational chemistry, measure in the laboratory, and attempt to observe in space.

Interestingly, fields such as drug and materials discovery face a similar problem; open source tools in the cheminformatics and machine learning space have transformed how novel molecules are discovered and/or designed [e.g. \citep{janet_machine_2020,kulik_making_2020,david_molecular_2020}]. By exploiting the scalability of chemically descriptive computer representations of molecules, we can systematically and exhaustively identify attractive candidates for astrochemical study. In this work, we demonstrate the feasibility and accuracy of such an approach on a well-characterized chemical inventory, the cyanopolyyne peak of the Taurus Molecular Cloud 1 complex (TMC-1). Combining unsupervised machine learning of chemical embeddings with ``classical'' supervised machine learning regressors, we are not only able to successfully reproduce observed molecular abundances, but recommend and predict the abundances of thousands of chemically related molecules. Perhaps most importantly, our approach does not require prior knowledge of the physical/chemical conditions, contrasting with conventional chemical modeling, which can rely on parameters that are not always known and/or cannot be directly determined. 

In this paper, we provide a verbose discussion of the workflow, theory, and implications of applying unsupervised machine learning for astrochemical inference; given the relatively niche intersection of cheminformatics, machine learning, and astrophysics, this paper is written with particular emphasis on the interpretation and reconciliation of machine learning predictions with chemical intuition. We begin by introducing machine representations of molecules, followed by details on the overall workflow and descriptions of various regressors. From there, we discuss and provide visualizations of the learned vector representations, and contextualizing them in the broader scopes of chemical inventories and networks, and finally discuss the use of these embeddings for recommendation and regression. 

\section{Computational methods}

\subsection{Molecule representation learning}

In order for quantitative comparisons to be made with machine learning methods, molecular features need to first be encoded into vector representations. A common approach is to hand pick features appropriate for the task at hand, for example the length of hydrocarbons or the number and types of functional groups, and express properties using additivity schemes \citep{benson_additivity_1958}. While these approaches are simple to implement, they are subject to the choice of features and as certain features may only occur in certain groups of molecules, hand picked features are neither scalable nor balanced in their approach to representations. Alternatively, more systematic [e.g. Coulomb matrices \citep{rupp_fast_2012}] and unsupervised approaches such as \textsc{mol2vec} \citep{jaeger_mol2vec_2018} provide means to generate molecule vectors without the need to choose descriptors. For this work, we have chosen the \textsc{mol2vec} approach, as it does not require molecular structures---instead, \textsc{mol2vec} repurposes the \textsc{word2vec} algorithm \citep{mikolov_efficient_2013} developed for natural language processing, and operates on linear string representations of molecules, specifically in the Simplified molecular-input line-entry system (SMILES) format \citep{weininger_smiles_1988,oboyle_towards_2012} commonly enumerated in large datasets.

The \textsc{mol2vec} algorithm decomposes the representation task into two aspects; unique atom environments defined by a radius hyperparameter are hashed with the Morgan algorithm \citep{morgan_generation_1965} to form a dictionary/corpus of substructures, which are subsequently used to train a multilayer perceptron---a continuous bag of words architecture \citep{mikolov_efficient_2013}---to learn a context-aware, unsupervised mapping of substructures (words) to molecule (sentences) vectors. Thus, for every canonical SMILES string that encodes every functional group and connectivity in a molecule, \textsc{mol2vec} generates $n$-dimensional vectors (in our case, 300-dimensions) as a sum of substructure vectors, capturing \emph{every} molecular feature. This model description of chemistry has been successfully used for a number of predictive tasks, for example drug activity screening \citep{das_repurposed_2021}, property prediction and rationalization \citep{zheng_identifying_2019}, and chemical space exploration and visualization \citep{shibayama_application_2020}.

By training the model on a diverse set of SMILES strings---as with many unsupervised approaches---the resulting embeddings become more holistic in their description of chemical properties. Developing a dataset for astrochemical purposes, however, requires striking a balance in descriptiveness and utility: cheminformatics datasets typically comprise large, biological molecules whereas those detected in the interstellar medium are smaller and oftentimes transient and unstable. Given that dataset bias is currently a well-known problem in word embeddings used for natural language applications \citep{bolukbasi_man_2016,basta_evaluating_2019}, we were mindful not to bias toward terrestrial chemistry albeit only qualitatively so. For this reason, we curated a comparatively small dataset of molecule SMILES balanced in its emphasis on molecules relevant to interstellar chemistry (KIDA, NASA PAH database), and small to medium sized molecules from several generalized datasets (QM9, ZINC, Pubchem, PCBA). For some datasets, SMILES notation for the molecules are not provided, for example the NASA PAH database and KIDA: the former provides cartesian coordinates, and the latter InCHI notation. In both cases, \textsc{OpenBabel} \citep{oboyle_open_2011} was used to convert these formats into SMILES strings. The sources of data are summarized in Table \ref{tab:dataset}, and correspond to a total of 6,883,279 entries which are filtered for duplicates, resulting in 3,316,454 unique canonical SMILES strings that constitute the training data for \textsc{mol2vec} and for molecule recommendations. While we have not performed a systematic analysis into the influence of each public dataset on the resulting embeddings, from Table \ref{tab:dataset}, it is easy to conclude that the number and diversity of molecules contained in astrochemical datasets like KIDA and the NASA PAH database alone would not be sufficient for an adequately descriptive embedding. We note also that this dataset comprises radicals and ions, however we have chosen not to include isotopologues although they can be readily encoded in SMILES strings.

\begin{table*}[ht]
    \centering
    \caption{Composition of the dataset used for this work; sources and number of entries within each dataset.}
    \begin{tabular}{l r c}
        \toprule
        Source & Number of entries & Reference \\
        \midrule
        ZINC & 3,862,980 & \citet{sterling_zinc_2015} \\
        PubChem A & 2,444,441 & \citet{kim_pubchem_2021} \\
        PCBA & 437,929 & \citet{wang_pubchems_2012} \\
        QM9 & 133,885 & \citet{ramakrishnan_quantum_2014} \\
        NASA PAHs & 3,139 & \citet{boersma_nasa_2014,bauschlicher_nasa_2018,mattioda_nasa_2020} \\
        KIDA & 578 & \citet{wakelam_2014_2015}\\
        TMC-1 & 87 & See Table \ref{tab:fulldata} \\
        \bottomrule
    \end{tabular}
    \label{tab:dataset}
\end{table*}

\subsection{Model pipeline}

\begin{figure*}
\centering
\includegraphics[width=0.8\textwidth]{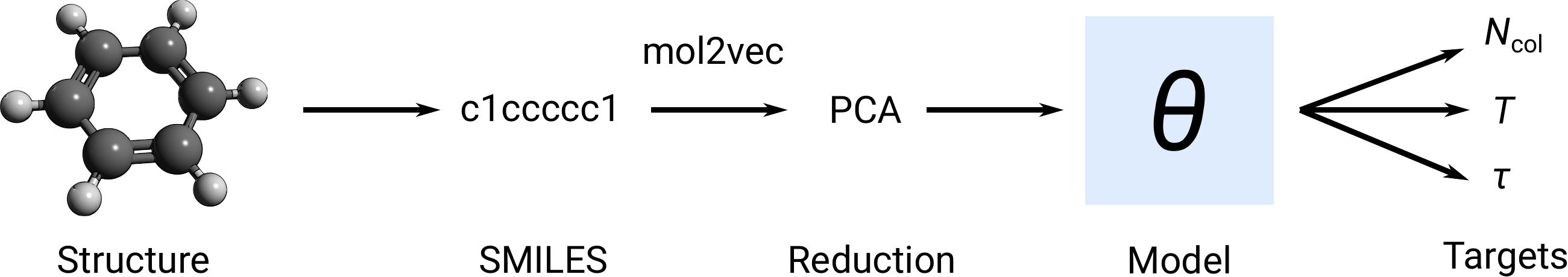}
\caption{Proposed workflow for unsupervised training and prediction of molecular properties. Molecular structures can be encoded in a number of ways, ranging from atomic Cartesian coordinates to internal coordinate (or Z-) matrices. Structures are standardized with canonical SMILES, ensuring uniqueness in the dataset, and transformed into molecule vectors using \textsc{mol2vec}. The dimensionality of the vectors are reduced via principal components analysis (PCA), and are used by various models to predict target properties.}
\label{fig:pipeline}
\end{figure*}

Figure \ref{fig:pipeline} illustrates the computational flow: molecular structures encoded as SMILES strings are passed to the trained \textsc{mol2vec} embedding model, generating 300-dimension feature vectors. Subsequently, we carry out a dimensionality reduction with principal components analysis (PCA) followed by clustering with $k$-means. Of the regression algorithms surveyed here, Gaussian Processes (GPs) are the most memory and compute intensive due to the factorization of large matrices, which for a naive implementation, scales with $O(n^3)$. Reduction with PCA decreases the memory usage substantially, and $k$-means clustering attempts to include only molecules that are of immediate relevance to those found in the astrophysical source under investigation. For details on the PCA dimensionality reduction, see Appendix \ref{sec:pca}. Prior to regression, we also perform feature standardization by scaling the feature values by their mean and variance, determined during training. In conjunction with the regressor specific regularization methods, standardization attempts to preserve sparsity in the model which in turn contributes towards mitigating overfitting and improving accuracy. The code---including software environment specification---can be found at \url{https://github.com/laserkelvin/umda}.

\subsection{Model specification and selection}

The final step of the pipeline shown in Figure \ref{fig:pipeline} is to perform supervised machine learning to predict column densities of molecules detected in TMC-1. To establish a baseline for performance, we tested a variety of commonly used machine learning algorithms, chosen for their simplicity, and some for their well-documented performance and interpretability. Table \ref{tab:model-list} organizes the methods used, and in Appendix \ref{sec:regressors} we provide a short overview of advantages and disadvantages of each. Each method is classified into whether there are learnable parameters or not; as we will discuss in subsequent sections, this will motivate the choice of method to apply.

The primary goal of each regressor is to accurately reproduce the observed column densities in TMC-1; for our purposes, this corresponds to the column densities of 87 molecules spanning from methylidyne \ce{CH} to cyanonaphthalene (c-\ce{C10H7CN}), with model accuracy measured by the mean squared error of the $\log_{10}$ column densities. To briefly summarize the data used for this work: 3.3 million molecules constitute the dataset for training the embedding and PCA model; 455,461 molecules form the subset of which are considered astrochemically relevant to TMC-1 through $k$-means clustering. For column density prediction, the data comprises 87 molecules with observed column densities toward TMC-1, bootstrapped (randomly sampled with replacement) with Gaussian noise ($\sigma=0.5$) added to the log column densities to yield an effective data set of 800 observations---see Appendix \ref{sec:bootstrap} for more details.

For all models except LR and BR, we perform hyperparameter optimization using grid search combined with 10-fold cross-validation, whereby the bootstrapped dataset is split into ten subsets of training and validation data, and 20\% of the species (17 molecules per split) are not used to fit the regressor.  Table \ref{tab:model-list} summarizes the hyperparameters that are tuned as part of the exhaustive grid search. All models used in this work are implemented in \textsc{scikit-learn} \citep{pedregosa_scikit-learn:_2011}. The hyperparameter and test scores for each model is summarized in Table \ref{tab:hparam_opt}. To assess the degree of overfitting, we perform learning curve analyses, whereby the trained model performance is evaluated as a function of dataset size based on 10-fold cross-validation; these details can be found in Appendix \ref{sec:bootstrap}.

\begin{table*}[ht]
    \centering
    \caption{Summary of models used in this work, their types, and respective optimized hyperparameters during cross-validation.}
    \begin{tabular}{l l l l}
        \toprule
        Model & Abbreviation & Category & Hyperparameter space \\
        \midrule
        Linear regression & LR & P & \\
        Ridge regression & RR & P & $\alpha$ \\
        Bayesian ridge regression & BR & P &  \\
        Support vector regression & SVR & P & $L_2$ and $\varepsilon$ regularization, $\gamma$ \\
        $k$-nearest neighbors & $k$NN & NP & Num. neighbors, distance metric \\
        Random forest & RFR & NP & Num. trees \\
        Gradient boosting & GBR & NP & Learning rate, num. estimators, subsample fraction, min. samples \\
        Gaussian process & GPR & NP & $\alpha$, kernel \\
        \bottomrule
    \end{tabular}
    \label{tab:model-list}
\end{table*}

\section{Results \& discussion}

\subsection{Vector representations of chemistry}

The first step in the proposed pipeline is the generation of vector representations of molecules via unsupervised machine learning using the \textsc{mol2vec} method, which in turn is adapted from the \textsc{word2vec} algorithm from the natural language processing domain. In this section, explore a few properties of the learned embedding including the possible manipulations and information compression.

\begin{figure}
    \centering
    \includegraphics{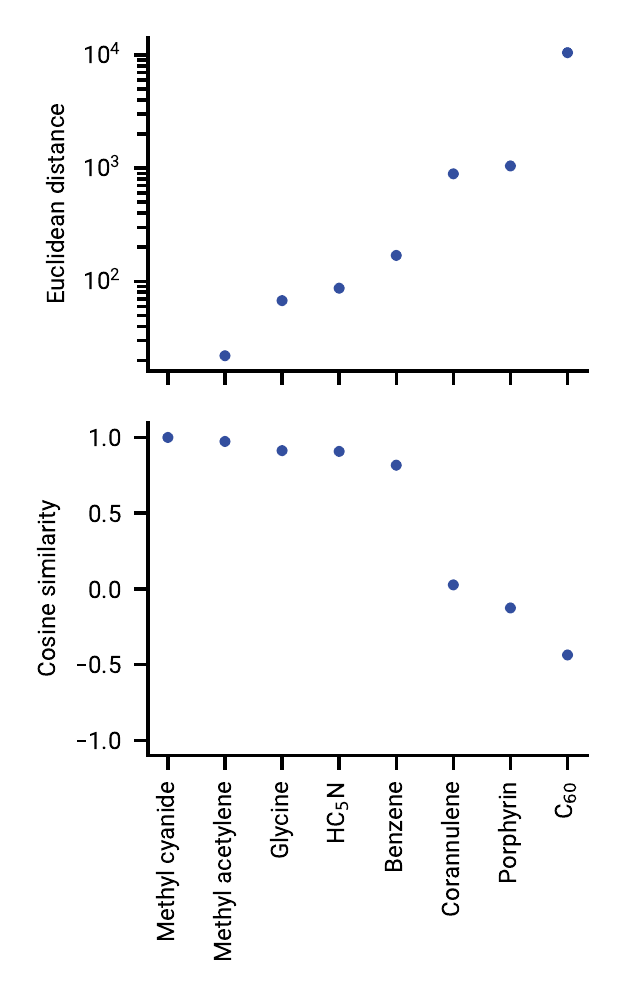}
    \caption{Similarity of arbitrarily chosen molecules with respect to methyl cyanide, given as a function of euclidean distance (top) in $\log$ space and cosine similarity (bottom).}
    \label{fig:cosine}
\end{figure}

To infer how chemical intuition is encoded in the \textsc{mol2vec} vectors, Figure \ref{fig:cosine} shows how the similarity, or conversely distance, changes over chemical space defined as a spectrum between two extremes: small molecules like methyl cyanide (\ce{CH3CN}) and monolithic structures such as buckminsterfullerene (\ce{C60}). The two metrics, euclidean distance and cosine similarity, provide slightly different insight into how the vectors behave---the latter is scale invariant as it simply measures the alignment of two vectors, while the former is not. This is particularly important in differentiating between molecules that are highly similar; for example, the margin between \ce{CH3CN} and methyl acetylene (\ce{CH3CCH}), and glycine (\ce{NH2CH2COOH}). Intuitively, the two methyl chains should be much more similar/closer in distance than \ce{CH3CN} and \ce{NH2CH2COOH} in terms of molecule size and functionalization (i.e. the former have methyl groups). 

To better understand the learned representation as well as the chemical inventory of TMC-1, we applied the Uniform Manifold Approximation and Projection (UMAP) method \citep{mcinnes_umap_2020} to visualize the embeddings of molecules detected toward TMC-1. This approach attempts to learn an approximation to the manifold of the embedding space, and produces a mapping between the approximate manifold and a lower dimensional representation in an unsupervised fashion. For our purposes, the goal is to visualize the two-dimensional chemical space comprised by molecules in TMC-1, whilst preserving the topology of the PCA reduced 70-dimensional vectors. 

\begin{figure*}[ht]
    \centering
    \includegraphics[width=\textwidth]{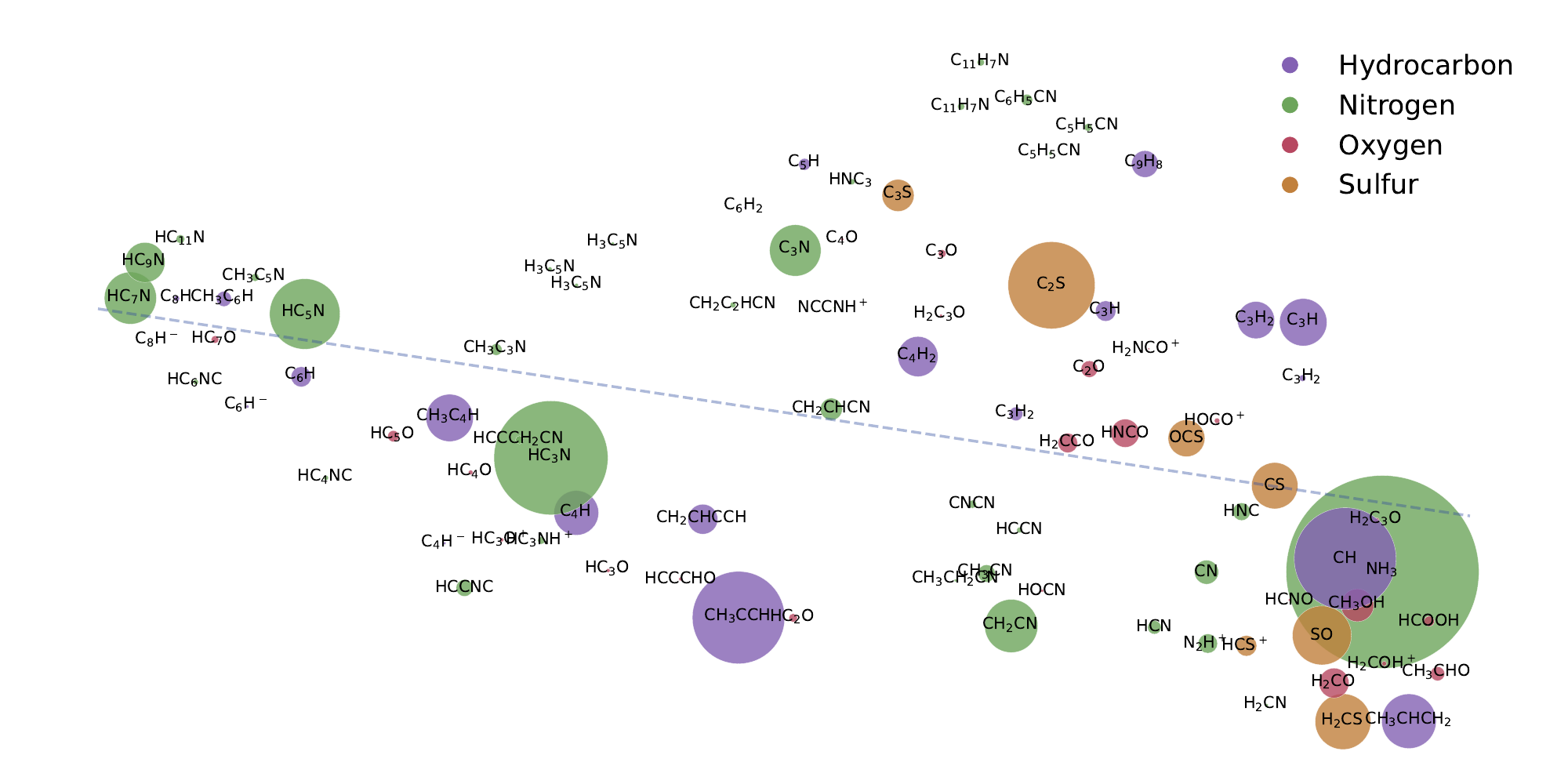}
    \caption{Visualization of the TMC-1 inventory, where the axes represent the UMAP learned two-component projection of the 70 dimensional vector molecule representation. Colors represent arbitrary classification of molecules, and were not used for UMAP training. The size of each scatter point corresponds to the molecular column density. The dashed line corresponds to a linear fit to the projection as a visual guide.}
    \label{fig:umapviz}
\end{figure*}

As shown in Figure \ref{fig:umapviz}, the UMAP method provides a unique perspective on chemical inventories, and validates some assertions of what is contained in the embeddings. For example, chemically similar molecules such as the cyanopolyynes and their methylated variants are located in the same region (left side of Figure \ref{fig:umapviz}), and trends within these families are also observable (i.e. chain elongation). Near the center-right of Figure \ref{fig:umapviz} we see clusters of smaller species, which constitutes the other extreme of molecules detected toward TMC-1, contrasting the large carbon chains. This dichotomy illustrates how chemical inventory characterization to date has largely followed a linear progression in chemical space---from right to left, as shown by the dashed line. The recent detections of large aromatic species such as indene [\ce{C9H8}, \citep{burkhardt_discovery_2021}] and cyanonaphthalenes [\ce{C11H7N}, \citep{mcguire_detection_2021}] intuitively correspond to a different type of chemistry, and is indeed identified in the UMAP projection as a cluster of molecules as somewhat orthogonal to the rest of detections, progressing from c-\ce{C3H2} through to \ce{C11H7N} towards the top.

\begin{figure}
    \centering
    \includegraphics[width=\columnwidth]{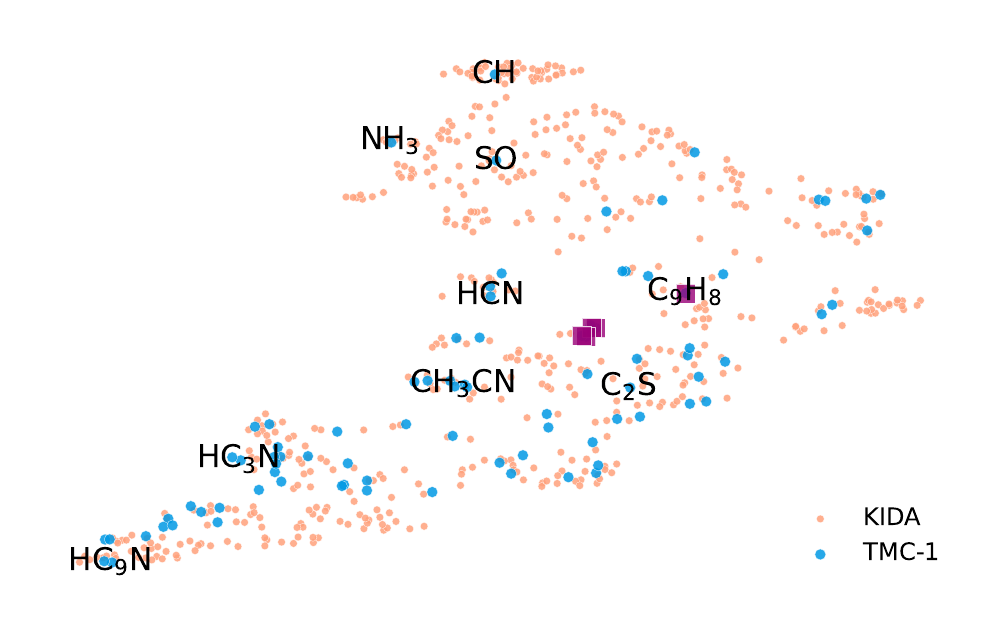}
    \caption{UMAP projections conditioned on the combined TMC-1 and KIDA molecule embeddings. Points with the shaded squares near the center of the image correspond to the aromatic ring molecules detected in TMC-1. Several molecules are annotated to represent the local region of chemical space.}
    \label{fig:kidaviz}
\end{figure}

Taking the inventory of TMC-1 into a broader context, Figure \ref{fig:kidaviz} shows a UMAP projection of molecules detected in TMC-1 and species contained in the KIDA network. The main observations here are that by in large, the KIDA network overlaps well with the inventory of TMC-1, corroborating with the current intuition of which species are important in the description of chemistry in dark molecular clouds. Where the KIDA points in chemical space are sparse, however, pertain to the recently detected aromatic molecules, which are clustered toward the center of Figure \ref{fig:kidaviz}. This naturally indicates that, for the proper description of aromatic chemistry, a network such as KIDA would need to increase coverage of aromatic species, to ``fill in the gap'' in the embedding space.

\subsection{Targets for astrochemical study \label{sec:targets}}

One aspect in astrochemistry that is currently poorly defined is the identification of potential molecules of interest for laboratory, observational, and/or modeling efforts. Using the chemical embeddings, likely targets for study can be readily identified simply by proximity in the latent space: molecules discovered in a given source or survey can be used as cluster centers, and molecules from the full dataset within an arbitrary distance threshold can be proposed for study in a systematic fashion. The basis for this is chemical similarity, in a way similar to how isomers and conformers of detected species are viable targets; we have simply vectorized this process.

Here, we provide recommendations for viable candidates of targeted efforts by selecting 100 nearest neighbors for each of the 87 non-isotopologue species detected in TMC-1, and using quantum chemistry to estimate their rotational constants and dipole moments as to assist in their discovery. The list is filtered for: (1) duplicates; (2) heavy elements outside of C, N, O, Si, P, S; (3) van der Waals complexes, leading to 1510 unique molecules. From this list, cartesian coordinates for each structure were generated and optimized with the UFF force field \cite{rappe_uff_1992} implemented in \textsc{OpenBabel} \cite{oboyle_open_2011}. The generated 3D structures are then refined at the $\omega$B97X-D/6-31+G(d) level of theory \citep{chai_long-range_2008,rassolov_6-31g*_1998}, chosen as a suitable compromise between computational expense and accuracy, in addition to well-known uncertainties and scaling factors \citep{lee_bayesian_2020}. For open-shell species, an unrestricted reference was used; results for these molecules should be taken conservatively---here we assume that the performance of the electronic structure method and basis provides the same systematic errors in the predicted structure as for the closed-shell case. It is likely that these species will require substantially more sophisticated treatments for desirable accuracy, including estimation of their fine structure properties. Geometry optimization was performed using the \textsc{geomeTRIC} package \citep{wang_geometry_2016} with gradients calculated using \textsc{Psi4} \citep{parrish_psi4_2017}. Out of the 1510 molecules, 148 were non-convergent either at the self-consistent field or geometry optimization steps.

The full list of molecules can be found in the Zenodo repository \dataset[10.5281/zenodo.5080543]{\doi{10.5281/zenodo.5080543}}; for brevity, we highlight and rationalize a few recommendations. Some general observations of the candidates include:

\begin{enumerate}
    \item The majority of molecules are unsaturated, containing at least one double or triple bond (1183 molecules, or 78\%);
    \item The vast majority of candidates are not pure hydrocarbons, with an average of at least 1.3 heteroatoms (338 pure hydrocarbons, 12\%);
    \item Most contain nitrogen, particularly as a \ce{-C#N} group (508 cyanides, 33\%);
    \item Relevant to aromatic/ring species, molecules with up to three rings are recommended. A substantial number of recommended species (436, 29\%) contain at least one ring. On a similarity basis, molecules with the same number of rings are typically recommended, although molecules with more or fewer rings are also identified in the search.
\end{enumerate}

These observations reflect the state of the dataset and the molecules currently detected in TMC-1: the majority of the molecules detected are indeed highly unsaturated, most are tagged with cyanides, and to date, only six aromatic species have been detected. Given the fact that a nearest neighbors approach was used to identify potential candidates, it is not surprising that the recommendations closely resemble those already detected. 

\begin{figure}
    \centering
    \includegraphics[width=\columnwidth]{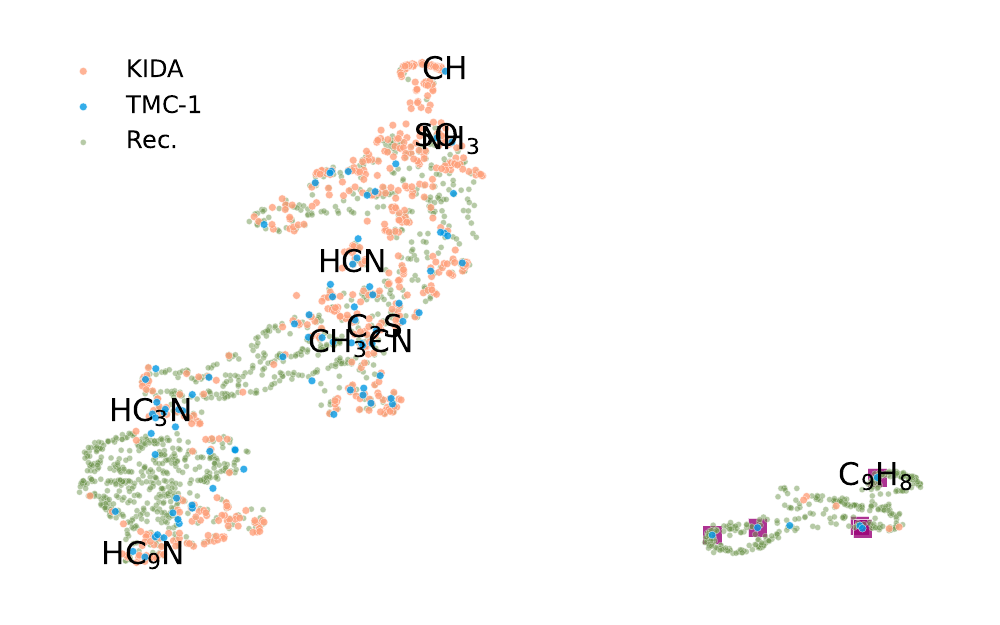}
    \caption{UMAP projection of the combined TMC-1 inventory (blue), KIDA network (peach), and the recommended molecules for study (green). Squares denote aromatic molecules detected in TMC-1.}
    \label{fig:umaprecs}
\end{figure}

To understand how the recommended molecules using this nearest-neighbors approach is complimentary to chemical inventories and model networks like KIDA, Figure \ref{fig:umaprecs} provides another UMAP learned visualization, trained on the TMC-1 detections, KIDA species, and recommendations. In the case of the former, the group of molecules to the left corresponds to the aromatic molecules detected in TMC-1, with a large number of new recommendations contributing \emph{significantly} toward three subclusters comprising \ce{C9H8}, \ce{C11H7N}, and \ce{C5H5CN}/\ce{C6H5CN} respectively. For KIDA, we see that the recommended species act to fill in large gaps in chemical space, particularly in the regions concerning larger species---between \ce{HC3N} and \ce{HC9N}. In both instances, the main observation is that the recommended molecules add to regions in chemical space that were previously sparse, particularly towards larger molecules. Viewed in this way, there are substantially fewer recommendations for smaller molecules as the inventory is relatively complete, compared to the number of isomers and conformers possible for larger species.

\subsection{Machine learning of chemical inventories}

From the prior sections, it is clear that our embedding successfully captures aspects of chemical intuition. The goal now is to relate the chemical features with physical parameters---the goal of chemical modeling---in a way that links molecules to directly and non-directly observable aspects of the interstellar medium. Here, we demonstrate the approach for column densities.

\begin{figure*}
    \centering
    \includegraphics{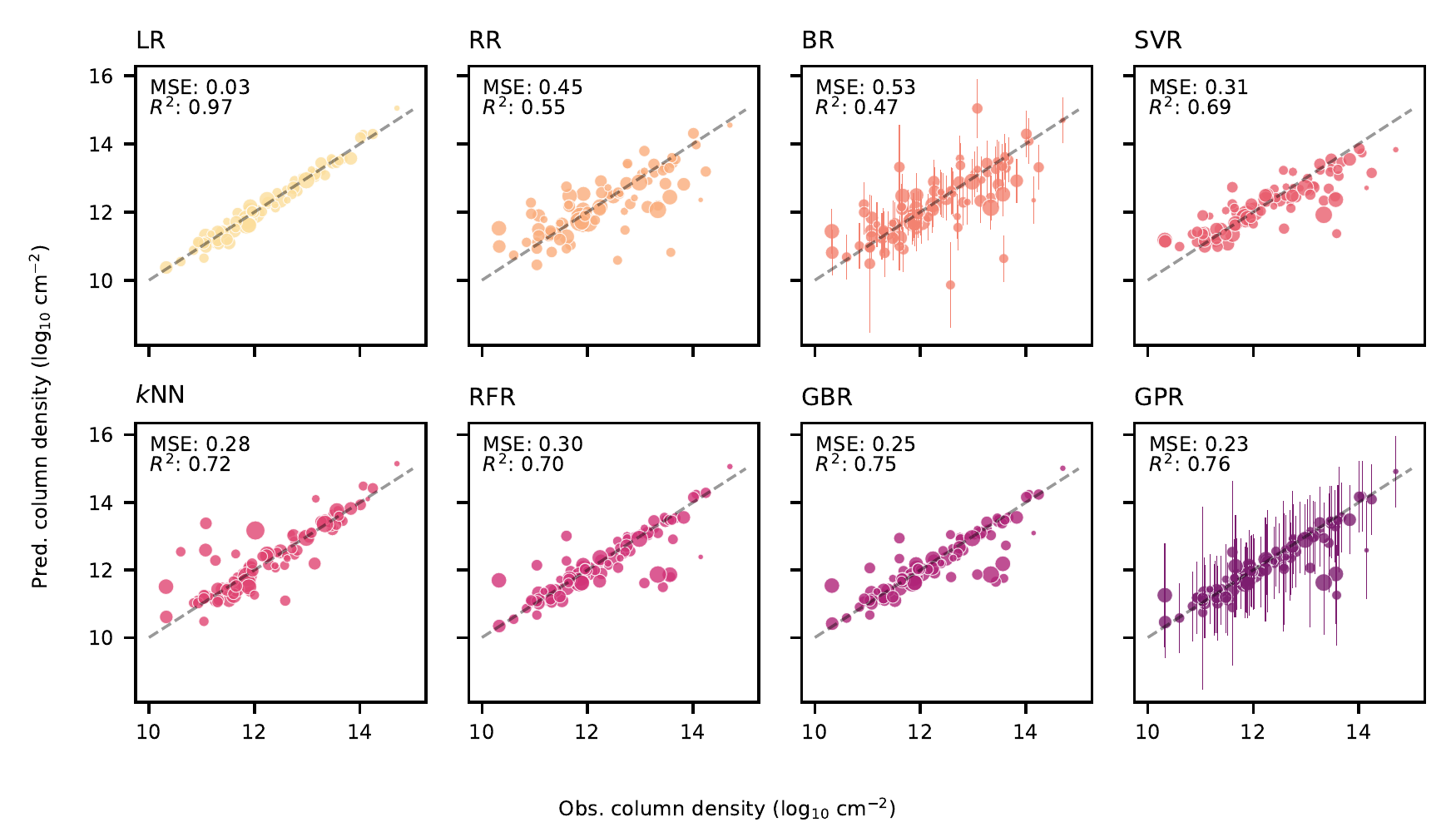}
    \caption{Observed column densities plotted against the corresponding model predictions. The size of each point is proportional to the molecular weight, as an approximate measure for molecular complexity. Three molecules that are underfit by several thousand orders of magnitude by the LR model are excluded from the plot and from the metric calculation (see text). For probablistic models, $1\sigma$ uncertainty is given as error bars.}
    \label{fig:r2plot}
\end{figure*}

Figure \ref{fig:r2plot} compares the true observed column densities with model predictions following model selection. We see that most algorithms---with the exception of linear regression, which drastically overfits without regularization---are able to reproduce the observed column densities remarkably well. In the case of linear regression, the coefficients become unrealistically large, and severely underfits three molecules (vinyl cyanide, thioformaldhyde, and butatrienylidene) by several thousand orders of magnitude (these are excluded from Figure \ref{fig:r2plot}), highlighting a significant need for heavily regularized models owing to the relatively small datasets and chemically simple molecules relevant to astrochemistry. 

The ridge regression models (frequentist and Bayesian) provide the necessary regularization to linear models, whilst performing qualitatively well and provides evidence that, at least locally, the abundance is approximately linear in the chemical space comprised by molecules detected in TMC-1, and that there are no specific \textit{classes} of molecules that demonstrate peculiarities in their abundance. This observation is reinforced by the fact that $k$NN performs extremely well even with only a few neighbors (Table \ref{tab:hparam_opt}). Were the opposite true, there would be systematic effects in the residuals, however the errors in the linear models appear normally distributed. As each molecule is or can be represented in the same embedding basis---regardless of whether they are ions, radicals, pure hydrocarbon or not---the linear function should be able to readily interpolate between molecules detected in TMC-1, and extrapolate to molecules not yet seen as simple extensions of the vector space they comprise. We note that in some sense, this is an intuitive result given the observed trends for hydrocarbon chains such as the cyanopolyynes and their methylated analogues, however here we generalize this trend beyond the one-dimensional slices in chemical space (i.e. the length of carbon chains). A short discussion regarding this aspect can be found in Appendix \ref{sec:complexity}.

From Figure \ref{fig:r2plot}, the bottom row comprises models with the lowest bias, namely $k$NN, RFR, GBR, and GPR. The case of $k$NN is particularly important, as it demonstrates that column densities can be expressed as smooth functions of local distance, and that learnable parameters are not necessary to describe this behavior. For the ensemble methods RFR and GBR, we see that both regressor types provide approximately the same degree of excellent performance---both methods are able to estimate feature importance, although in the current implementation we are unable to interpret the features directly as they are learned with unsupervised methods. However, future applications could correlate the embedding dimensions with hand picked features, which in turn could provide a means to translate which aspects of molecules are most critical to the chemistry of an environment.

The two probabilistic models, BR and GPR, warrant some extra discussion. In particular, BR provides a highly attractive approach to modeling chemical inventories, as it provides an extremely simple and regularized method to not only obtain abundances, but also uncertainties. For the vast majority of molecules, we are able to accurately reproduce the observed abundances within an order of magnitude, and certainly within $2\sigma$ uncertainty. In the case of GPR, we substantially increase the modelling flexibility, as well as having a significantly more interpretable prediction uncertainty: two outlier points are ascribed much larger uncertainties than others, one of which is consistently overpredicted by each regressor (\ce{NCCNH+}). This provides a reason to revisit the column density estimate by \citet{Ag_ndez_2015}, as to investigate whether there are anomalies in the chemistry or in the embeddings.

\begin{figure}
    \centering
    \includegraphics[width=\columnwidth]{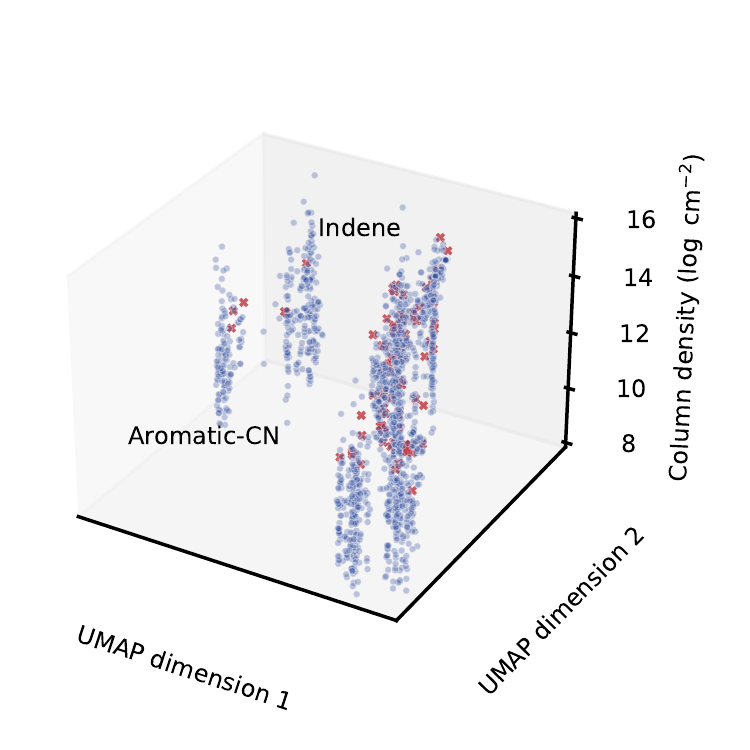}
    \caption{Scatter plot of the UMAP space for the 1510 recommended molecules (blue circles) and molecules detected toward TMC-1 (red crosses), and the corresponding GP predictions of column density. Locations of aromatic molecules are annotated.}
    \label{fig:reccolumns}
\end{figure}

Having established the dataset performance of each machine learning method, we can now use them for extrapolation, predicting column densities for unseen molecules. For the sake of brevity, we utilized the trained GPR model to predict column densities of the 1510 recommended molecules from Section \ref{sec:targets}. Figure \ref{fig:reccolumns} illustrates the result in three dimensions: the horizontal plane represents the UMAP learned 2D projection of the chemical space spanned by the TMC-1 dataset and the recommended molecules, and the vertical axis the GPR predicted column density. In the left cluster of points, which contain the majority of detected molecules, we see that a significant number of recommended species are predicted to have column densities ($10^{10-12}$\,cm$^{-2}$) comparable to those already detected---laboratory and computational efforts would likely enable their detection, or at least derivations of upper limits that can be used to refine machine learning and chemical models.

\subsection{Comparisons with chemical models}

\begin{figure}[ht]
    \centering
    \includegraphics[width=\columnwidth]{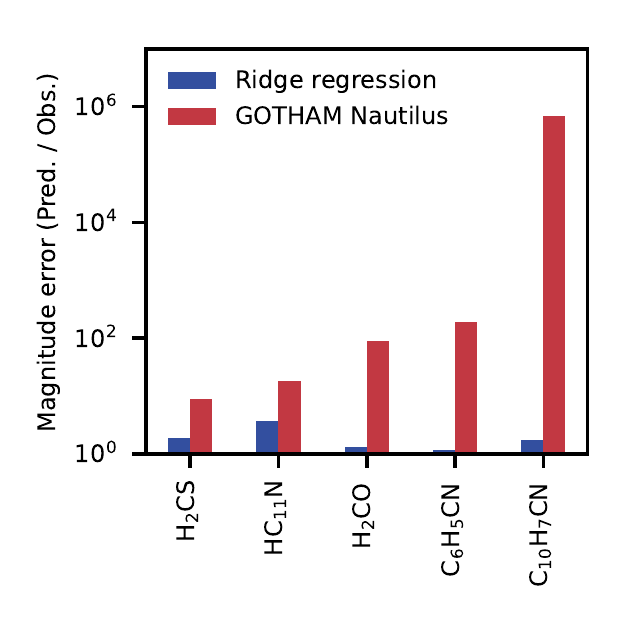}
    \caption{Comparison of molecular abundances predicted by the ridge regression model (blue) and the current state-of-the-art chemical model, \textsc{nautilus} \citep{ruaud_gas_2016}, for TMC-1 (red), given as an the magnitude for the predicted and observed ratio.}
    \label{fig:model_comp}
\end{figure}

In the preceding sections, we have evaluated the performance of various machine learning methods for predicting molecular column densities. A natural comparison to be made is with state-of-the-art kinetic chemical models of TMC-1, which have been the main method of choice for predicting abundances of molecules \textit{a priori}. Here, we utilize the three-phase chemical model \textsc{nautilus} \citep{ruaud_gas_2016} with the latest chemical network, elemental abundances, and physical conditions used to describe the formation of aromatics by \citet{burkhardt_discovery_2021}. Figure \ref{fig:model_comp} compares the predicted/observed ratios for five chosen molecules as samples across molecular complexity. Ridge regression is chosen as the baseline algorithm for comparison, and we see that it reproduces the abundance of each molecule within an order of magnitude of the observations. For chemical models the molecular abundance is time-dependent, and for the purposes of discussion we have chosen the time slice where the cyanopolyynes have their peak abundances. Under these conditions, the relatively simple species \ce{H2CO}, \ce{H2CS}, and \ce{HC11N} can be seen to agree with the observed abundances to within one to two orders of magnitude. For the aromatic rings, representing more complex species, the chemical model significantly under predicts their abundance; for cyanonaphthalene, this discrepancy is nearly six orders of magnitude, as originally discussed in \citet{mcguire_detection_2021}.

Beyond abundances, however, the current machine learning approach shares little to no similarities with chemical modeling. While Figure \ref{fig:model_comp} shows quite definitively that linear regression is able to reproduce observed abundances accurately, chemical models are an avenue for inferring physical information about an astrophysical source by reproducing abundances, albeit a complex and difficult process \citep{agundez_chemistry_2013,herbst_formation_1973,van_dishoeck_comprehensive_1986}. With each new interstellar molecule detection, the chemical network must be updated with new species hypothesized to be important in its formation and destruction, along with reaction rates measured, or more commonly, approximated \citep{wakelam_reaction_2010}. In contrast, the allure of the machine learning approaches is the ease of extension: new molecules can be added simply using SMILES strings, and as we have demonstrated in this work, can readily scale up to hundreds to millions of molecules. As we have alluded to in the previous section, a natural extension of this work is to connect physical parameter inference from chemical models with the generalizability of machine learning models: the latter informs the former by providing abundances and constraints on unobserved species, as well as providing recommendations for new molecules to add to chemical networks. Imputation through machine learning thereby results in a self-consistent and systematic approach to the astrochemical inference---we intend to explore and expand upon these ideas in forthcoming work. 

\subsection{Generalizations beyond TMC-1}

The approach we have detailed here essentially comprises two distinct machine learning parts: an unsupervised molecule embedding learning and compression task, followed by supervised training of regressors on molecular inventories. The former creates general purpose vectors in a latent chemical space, and is not limited to the study of chemical inventories nor regression---indeed, \textsc{mol2vec} vectors have been used for a range of machine learning tasks such as drug activity prediction. For the latter, chemical inventory regressors can be applied in two ways. First, the pre-trained TMC-1 model can be used to quantitatively assess chemical differences between sources (particularly dark molecular clouds) as a form of ``chemical baseline''. For example, the TMC-1 regressors can be used to identify systematic offsets (bias) and specific deviations (variance) in chemical abundances without re-training, which could be used to infer dynamics/kinematics unique to a given source, for example core collapse [e.g. L1521E \citep{hirota_l1521e_2002}] and protostar/warm carbon chain chemistry [e.g. L1527 \citep{sakai_detection_2007}].

The second application we foresee with machine learning regressors pertains to other well-characterized chemical inventories outside of TMC-1. Molecule rich, prototypical source such as VY Canis Majoris and IRC+10216 make excellent candidates for quantitative inventory analysis, albeit with significantly more complicated dynamics such as shock chemistry and photoinduced processes. For these applications, the simpler regressors reported here are unlikely to model considerations such as radial and angular extent or time-dependence; rather, more sophisticated parametric models such as neural networks will be required.

\section{Conclusions}

In this work, we have demonstrated the viability for simple machine learning models to learn and predict entire chemical inventories. Combining the \textsc{mol2vec} model embeddings with algorithms as simple as linear regression, we are able to reproduce the column densities of 87 molecules detected toward TMC-1 to well within an order of magnitude, without the need for prior knowledge pertaining to the physical conditions of the source. With this, we show that the molecule embeddings can be used to identify new likely candidates for interstellar detection and study, based on quantitative measures of chemical similarity between molecule vectors as in a nearest neighbors approach, and using the machine learning models to predict their expected column densities as one way to assess detectability. The attractiveness of our approach is the ability to \emph{systematically} infer the presence of astrophysically important molecules that are not directly observable, for example those without a rotational spectrum, or where conditions are unfavorable (e.g. partition functions), and to provide a baseline for determining the role of dynamical effects, such as grain-surface chemistry. In this way, predictions from machine learning models can be used to impute chemical networks used in conventional chemical modeling, from which we can confidently and comprehensively derive astrophysical insight. 

\acknowledgments

The authors thank the reviewers for insightful comments and feedback, in particular regarding model regularization and the suggestion to bootstrap the dataset. The National Radio Astronomy Observatory is a facility of the National Science Foundation operated under cooperative agreement by Associated Universities, Inc.  The Green Bank Observatory is a facility of the National Science Foundation operated under cooperative agreement by Associated Universities, Inc. J.P. and V.V. acknowledges funding and research support from the SAO REU program; the SAO REU program is funded in part by the National Science Foundation REU and Department of Defense ASSURE programs under NSF Grant no.\ AST-1852268, and by the Smithsonian Institution. M.C.M and K.L.K.L. acknowledge financial support from NSF grant AST-1908576 and NASA grant 80NSSC18K0396.

\newpage

\bibliography{references, molecules, brett}
\bibliographystyle{aasjournal}

\appendix

\renewcommand{\thefigure}{A\arabic{figure}}
\renewcommand{\thetable}{A\arabic{table}}
\renewcommand{\theequation}{A\arabic{equation}}
\setcounter{figure}{0}
\setcounter{table}{0}
\setcounter{equation}{0}

\section{Dimensionality reduction \label{sec:pca}}

\begin{figure}[ht]
    \centering
    \includegraphics{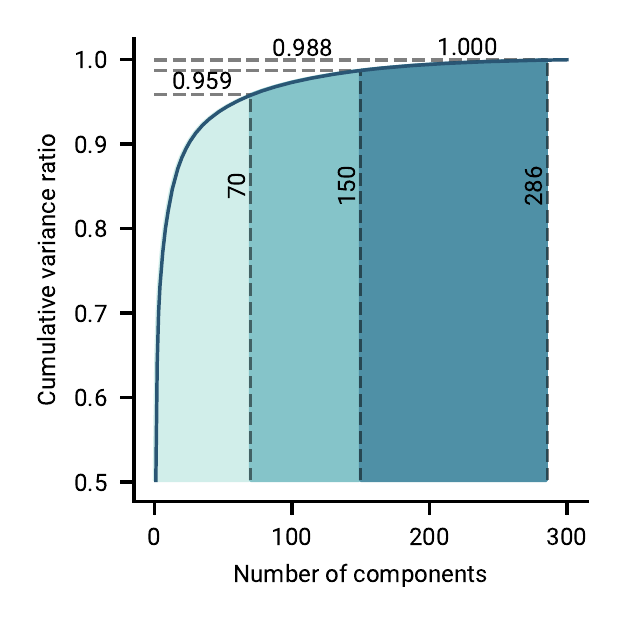}
    \caption{Cumulative explained variance ratio as a function of the number of components in the incremental PCA. The dashed lines represent the number of components that approximately correspond with $2\sigma$, $2.5\sigma$, and $3\sigma$ variation.}
    \label{fig:pca}
\end{figure}

The PCA model is trained on the full 3.3 million molecules to identify an adequate number of dimensions required to explain variation in chemical space, whilst providing computational benefits. Ultimately, we chose to use 70 principal components, corresponding to a 0.96 explained variance ratio (Fig. \ref{fig:pca}), or just over $2\sigma$ of variation accounted for by the components.

\section{Supervised machine learning regressors \label{sec:regressors}}

The simplest parametric and non-parametric models considered in this work are linear regression and $k$-nearest neighbors ($k$NN) respectively, which represent the two types of model abstraction. The former makes the assumption that the column density varies linearly on a \emph{global} scale in chemical space, while the latter expresses the column density as a distance weighted function. Thus, LR parameterizes a function that dictates the abundance of molecules decreases linearly from small to large molecules (i.e. chemical complexity), whereas for $k$NN the focus is \emph{local} chemical similarity.

As we have seen in the results, however, linear regression results in significant overfitting, whereby the linear coefficients become extremely large in order to fit the majority of the dataset. To alleviate this, ridge regression includes an $L_2$ coefficient norm penalty to the loss function, which encourages the linear coefficients to be small. A variation of this is a probabilistic treatment of the model coefficients, defined in Bayesian ridge regression, where predictions are linear coefficients sampled from gamma distributions, whose parameters are determined by maximizing the log marginal likelihood during fitting.

The remaining models are more sophisticated in their use of embedding space. First, $\varepsilon$-support vector regression [SVR; \citep{drucker_support_1996,platt_probabilistic_1999}] builds on top of linear regression by applying an $\varepsilon$ regularization term and a kernel transformation to the features prior to regression: in doing so, the model is capable of capturing nonlinearity in the embeddings while maintaining a simple linear mapping onto the column densities. For this work, we consider a radial basis function kernel with width $\gamma$. The ensemble learning methods we employ include random forest [RFR, \citep{liaw_classification_2002}] and gradient boosting [GBR, \citep{friedman_stochastic_2002}]; these often substantially improve upon linear models in both bias (boosting) and variance (forests) performance by aggregating the results of multiple weak models that collectively form an ensemble. The former comprises submodels that are based on randomly selected features with replacement, with the result given as a error weighted average of all submodels predictions. The latter sequentially trains predictors with weighted data, where the weights are given by the gradient of the error from the prior predictor; a large number of estimators thus tend to yield results with very low bias, and as an ensemble, low variance.

The last method we consider are GPs, which treat \emph{functions} of molecular properties as a stochastic process collectively defined by mean and covariance/kernel functions which---similar to $k$NN---expresses the column density as a function of distance in the embedding space. Among the regression models considered here, GPs are unique in their probabilistic nature and being the most flexible, given the ability to design kernel functions that optimally suit the embedding space. Given, however, that the embeddings are not directly interpretable, we provide here only a simple mixture kernel comprising three subkernels: the sum of rational quadratic, dot product, and white noise kernels. This covariance function was formulated assuming two components: the rational quadratic kernel describes short ranged, smoothly variability in the latent space, while the dot product kernel explicitly models pairwise, linear contributions between molecules. The former dominates for molecules that are chemically very similar (e.g. \ce{HC5N} and \ce{HC7N}), and the latter contributes for molecules that are at different scales of chemical complexity or size, however share common features such as functional groups (e.g. \ce{HC5N} and benzonitrile \ce{C6H5CN}). Finally, the white noise kernel provides modeling flexibility in describing uncertainty/noise in the observed column densities.

For each of the models, we perform hyperparameter tuning using grid search with cross-validation; the optimal hyperparameters are shown in Table \ref{tab:hparam_opt}.

\begin{table}[]
    \centering
    \caption{Tuned hyperparameters based on grid search with 10-fold cross-validation. For all other omitted hyperparameter values, the default values in \textsc{scikit-learn} were used.}
    \begin{tabular}{l l l l}
    \toprule
        Method & Hyperparameter & Value & Validation MSE\tablenotemark{a}  \\
    \midrule
        LR & & & $2\times10^21$ \\
        RR & $\alpha$\tablenotemark{b} & 1. & 0.54 \\
        BR & $\alpha$\tablenotemark{b} & 1.7 & 0.51 \\
        SVR & $C$ & 100 & 0.48 \\
        ~ & $\varepsilon$ & $10^{-1}$ & 0.58 \\
        $k$NN & Distance & cosine & 0.22 \\
        ~ & $N$ & 30 & \\
        RFR & $N$ & 50 & 0.53 \\
        GBR & Learning rate & 0.01 & 0.48 \\
        ~ & Samples per leaf & 0.3 & \\
        ~ & Samples per split & 0.1 & \\
        ~ & Number of estimators & 100 & \\
        ~ & Subsampling fraction & 0.8 & \\
        GPR & $\alpha$ \tablenotemark{c} & $3\times10^{-5}$ & 0.28 \\
    \bottomrule
    \end{tabular}
    \tablenotetext{a}{Mean squared error based on the non-bootstrapped observations.}
    \tablenotetext{b}{$L_2$ penalty term.}
    \tablenotetext{c}{Noise added to the diagonal of the kernel matrix for numerical stability.}
    \label{tab:hparam_opt}
\end{table}

\section{Dataset bootstrapping \label{sec:bootstrap}}

\begin{figure}
    \centering
    \includegraphics{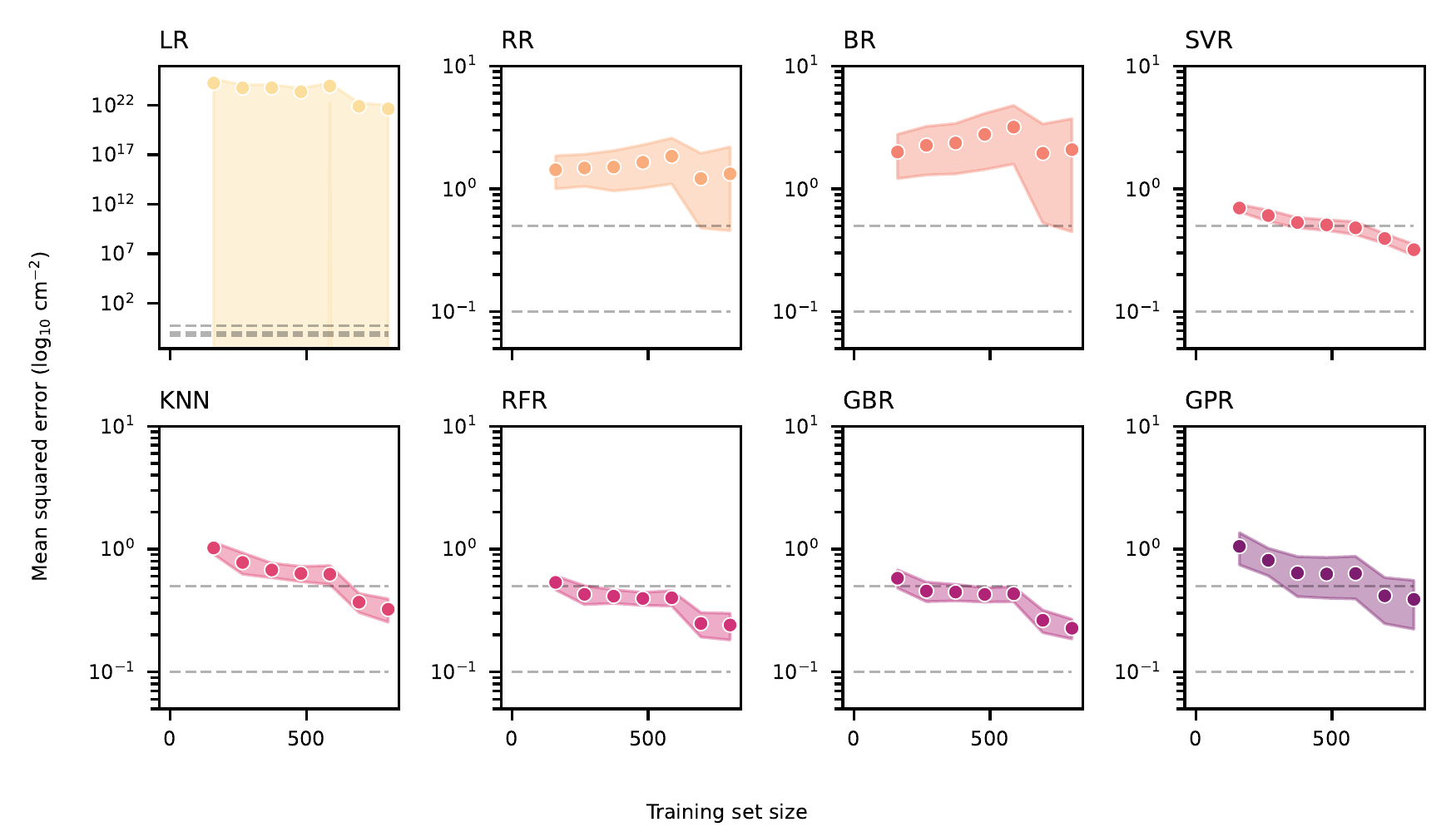}
    \caption{Learning curve analysis using the bootstrapped dataset for each supervised regressor. The abscissa corresponds to the number of training examples, while the ordinate represents the mean squared error based on the non-bootstrapped, true column density values. The shaded regions represent $1\sigma$ variation in the error, estimated with 10-fold cross-validation.}
    \label{fig:learningcurve}
\end{figure}

One aspect that became apparent over the course of this work was the susceptibility of supervised regressors overfitting the TMC-1 observations. Given that the PCA vectors constitute 70 dimensions, and that the true dataset only constitutes 87 observations, even linear models can be considered as overparameterized. To alleviate this, we used the bootstrap method to effectively generate ``new data'', whereby the original dataset is resampled and Gaussian noise ($\sigma=0.5$) is added to the log column densities, yielding an effective dataset of 800 points. As seen in Figure \ref{fig:learningcurve}, more training examples improve the performance of each model, as measured by the mean squared error with respect to the observed TMC-1 column densities.

\section{Comparison of hand picked descriptors and unsupervised embeddings}

In this work, we utilize molecule embeddings that are learned via unsupervised machine learning. While the main allure of this approach is a scalable method for featurization, it is important to compare with handpicked features of molecules for supervised regression. To make this comparison, we arbitrarily chose 15 molecule descriptors to constitute the feature vectors to perform regression as was done with the \textsc{mol2vec} vectors. The descriptors we chose are implemented in RDKit \citep{landrum_rdkit_2020}, which include: number of atoms, number of bonds, molecular weight, average bond order, number of aromatic rings, number of valence electrons, \textit{FpDensityMorgan}(1, 2, 3), and the number of atoms for carbon, oxygen, nitrogen, sulfur, and phosphorus. The descriptors should comprise what is necessary to describe the various radicals, charge states, and molecule complexity associated with molecules in TMC-1. Unlike the ``production'' workflow, we did not perform feature normalization, and ridge regression was used to make the comparison; both choices were intended to establish a baseline in the expected performance, and ridge regression was chosen given to its relative model simplicity and its $L_2$ regularization term to prevent overfitting.

\begin{figure}
    \centering
    \includegraphics{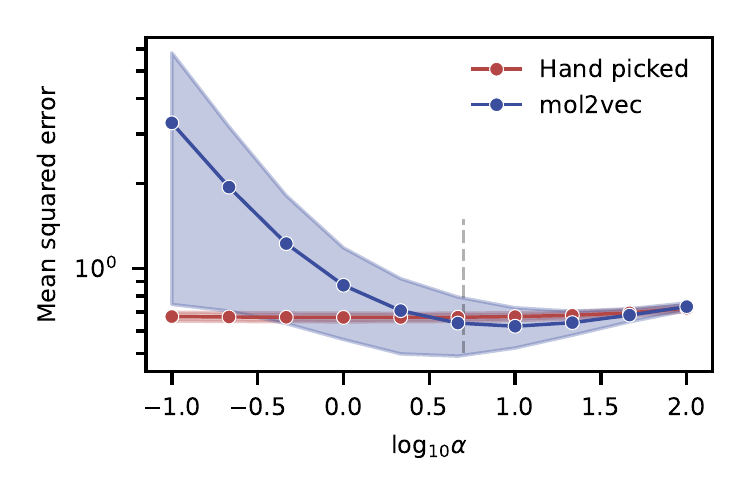}
    \caption{Performance of ridge regressors using hand picked (red) and \textsc{mol2vec} (blue) vectors, as a function of regularization term $\alpha$. The dashed line represents the value of $\alpha$ used in Figure \ref{fig:handpick_lc}. Increasing regularization strength goes from left to right. The shaded region corresponds to $1\sigma$ based on 20-fold cross-validation}
    \label{fig:handpick_alpha}
\end{figure}

Figure \ref{fig:handpick_alpha} compares the training and test errors for ridge regressors using hand picked and \textsc{mol2vec} vectors, based on a 0.8/0.2 train/test split of the bootstrapped dataset. We see that for very small values of $\alpha$---corresponding to minute regularization---the \textsc{mol2vec} vectors tends to overfit and yield a large error with respect to the true column densities. In the range of $\alpha = $1--30, the \textsc{mol2vec} model yields better performance than the hand picked alternative, and at larger values of $\alpha$ both models begin to be over-regularized. The regression performance for hand picked features stays relatively constant for the range of $\alpha$ values used, in comparison to the results seen for the \textsc{mol2vec} features, which shows a large degree of variability.

\begin{figure}
    \centering
    \includegraphics{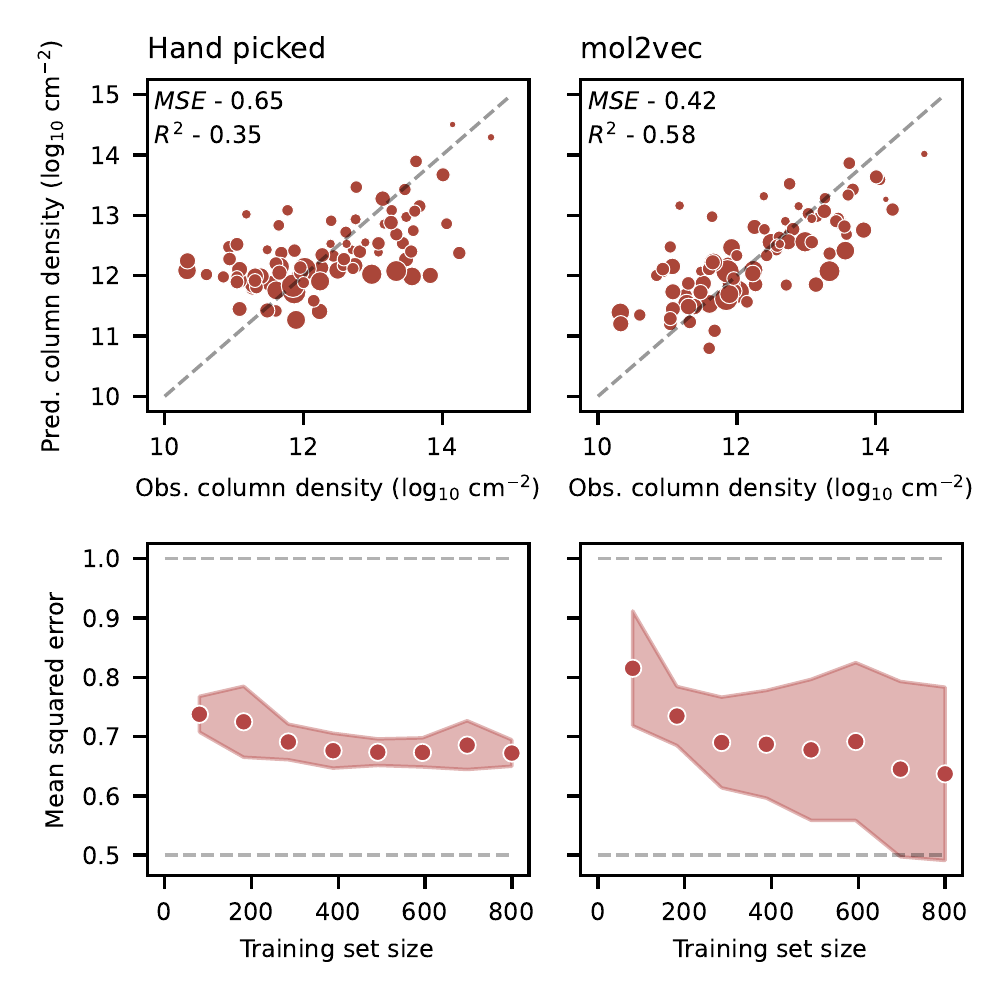}
    \caption{Comparison of $R^2$ plots (top row) and learning curves (bottom row) between hand picked features (left column) and \textsc{mol2vec} embeddings (right column). Shaded regions in the learning curves represent the $\pm1\sigma$ in model performance, as estimated with 20-fold cross-validation.}
    \label{fig:handpick_lc}
\end{figure}

As an alternative visualization of model performance, Figure \ref{fig:handpick_lc} (top row) shows the prediction errors and corresponding $R^2$ values (measured with respect to the true column densities) based on both featurization methods when fit to the full bootstrapped dataset, with $\alpha = 5$. The motivation here is to test the performance of features, given the same nominal modeling flexibility and to mitigate variance across cross-validation sets. We see that the hand picked features display a significant degree of bias (i.e.  in the predictions owing to the fact that many molecules are not adequately described). In contrast, the \textsc{mol2vec} has significantly less bias and variance, seen both in the scatter as well as the mean squared error and $R^2$ metrics.

\begin{figure}
    \centering
    \includegraphics{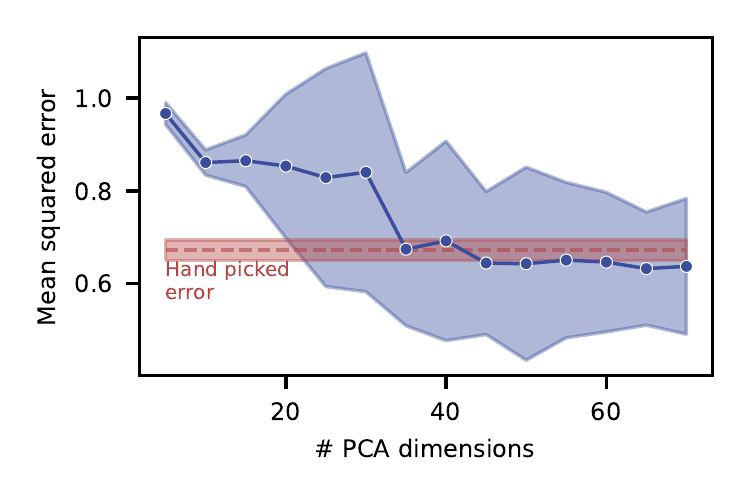}
    \caption{Performance of ridge regression models ($\alpha = 1$) using the \textsc{mol2vec} embeddings, as a function of the number of PCA dimensions used for regression. The red and blue traces represent hand picked and \textsc{mol2vec} representations respectively; the mean error is represented by the dashed line for the former, and scatter points for the latter. The shaded regions correspond to $1\sigma$ variation determined through 10-fold cross-validation.}
    \label{fig:handpicked_pca}
\end{figure}

Finally, Figure \ref{fig:handpicked_pca} shows the performance of a ridge regression model with an amount of regularization ($\alpha = 5$) where both models perform well, as a function of the number of PCA feature dimensions in the \textsc{mol2vec} embeddings, compared to hand picked features as an effective baseline. We see that for a comparable number of dimensions to the hand picked features (${\sim}$15--20 dimensions), the \textsc{mol2vec} features actually result in poorer performance: this is not surprising as the PCA model was trained on full 3.3 million molecules, which substantially biases the components towards a general description of chemistry not necessarily relevant to that of TMC-1, in particular highly reactive species do not constitute a significant part of typical public databases. For specific tasks, such as comparisons between dark molecular clouds, one could perform the PCA on the TMC-1 dataset specifically to obtain the transformations that best describe dark cloud chemistry, which would likely then decrease the number of dimensions required for better model performance. Nonetheless, this result makes the distinction that for a small number of dimensions, features chosen by human intuition can potentially provide better performance than embeddings obtained via unsupervised learning.

\section{Abundance and molecular complexity \label{sec:complexity}}

\begin{figure}
    \centering
    \includegraphics[width=\textwidth]{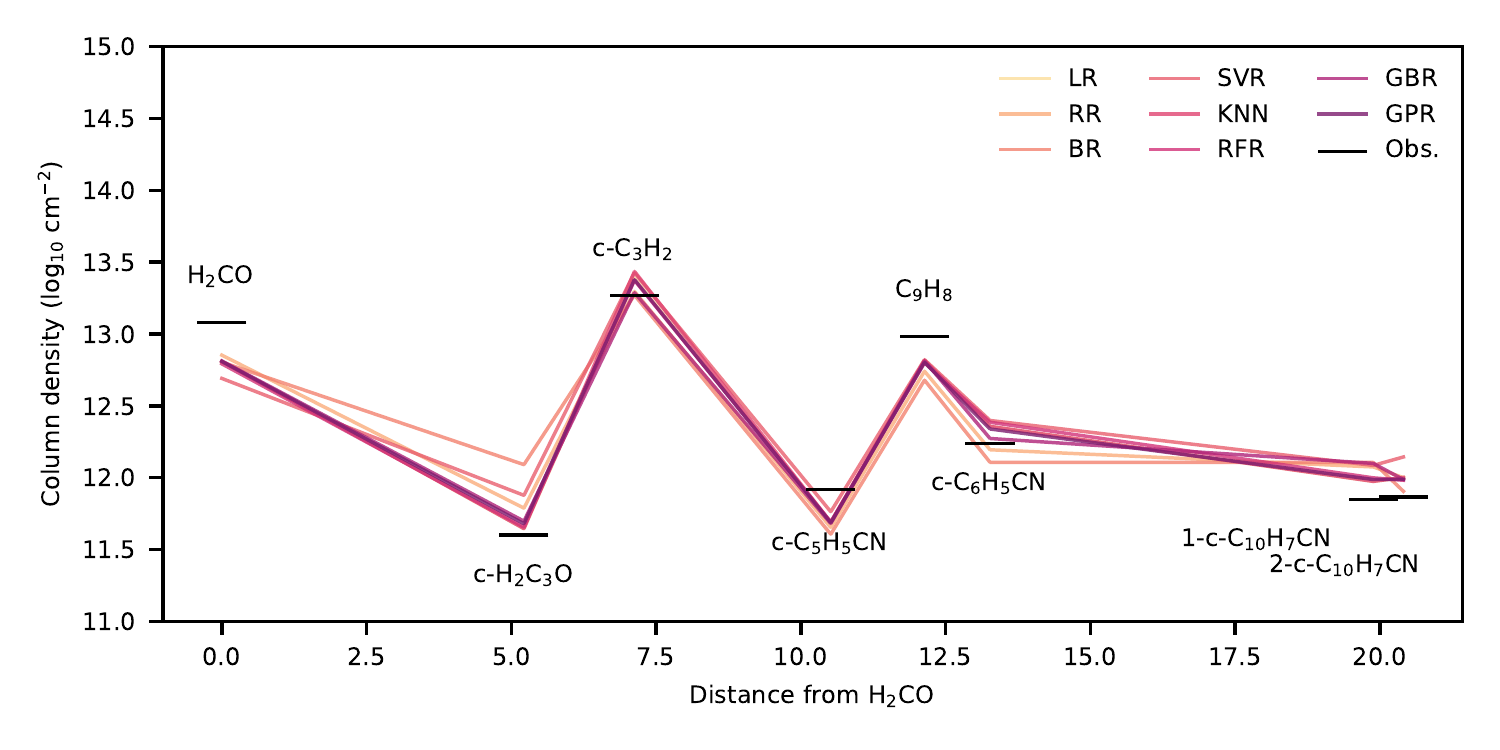}
    \caption{Predicted column densities for select molecules (ordinate) versus euclidean distance from \ce{H2CO} (abscissa). The black lines represent observed column densities.}
    \label{fig:demoset}
\end{figure}

To more quantitatively assess model performance as a function of molecular complexity, we can perform linear interpolation between two arbitrary molecules as one-dimensional slices in chemical space, and look for systematic errors in the predicted column densities. Figure \ref{fig:demoset} interpolates between formaldehyde (\ce{H2CO}) and 2-cyanonaphthalene (c-\ce{C10H7CN}) and identifies the nearest detected molecule to each interpolated point, with the distance from \ce{H2CO} providing a naive, relative representation of chemical complexity---the larger the distance, the more ``complex'' the molecule is. From a modeling perspective, it is important to highlight that there is no individual metric that best describes complexity, in a way that reflects the common intuition that abundance is anticorrelated with molecular complexity. Figure \ref{fig:demoset} highlights this well, as the number of atoms---which is partially encoded in the distance, and a common metric used in astrochemistry---barely correlates with column density. However, we see that all methods are able to predict the abundance of molecules within this small subset to well within an order of magnitude, and in particular, the linear models show that the abundance can be expressed as linear functions of 70 PCA dimensions, even if the mapping between individual descriptors and the abundance is not.

\section{Molecule dataset}

Table \ref{tab:fulldata} summarizes the molecules detected toward TMC-1 and their respective references, totaling 87 unique species. Based on our current knowledge of the chemical inventory of TMC-1, the size and complexity of this structure---and on grounds of chemical similarity with the remaining seven---would intuitively lead to a vanishingly small ``true'' column density.

\startlongtable
\begin{deluxetable}{lllc}
\tablecolumns{4}
\tablewidth{0pt}
\tablecaption{Regression dataset comprising molecules detected toward TMC-1. The references provided correspond to the source used for the column density, not initial detection. \label{tab:fulldata}}
\tablehead{\colhead{Formula} & \colhead{SMILES} & \colhead{Column density} & \colhead{Reference} \\ \colhead{~} & \colhead{~} & \colhead{($\log_{10}$ cm$^{-2}$)} & \colhead{~}}
\startdata
 \ce{CH3C6H} &             CC\#CC\#CC\#C &                 12.4914 &    \citet{Remijan_2006} \\
  \ce{CH3C4H} &                 CC\#CC\#C &                 13.4771 &    \citet{MacLeod_1984} \\
  \ce{CH3C5N} &             CC\#CC\#CC\#N &                 11.9243 &    \citet{Remijan_2006} \\
  \ce{CH3C3N} &                 CC\#CC\#N &                 12.2553 &    \citet{Remijan_2006} \\
    \ce{N2H+} &                  N\#[NH+] &                 12.6990 &       \citet{Choi_2017} \\
     \ce{NH3} &                         N &                 14.6998 &    \citet{Gratier_2016} \\
  \ce{CH3OH} &                        CO &                 13.1614 &    \citet{Gratier_2016} \\
     \ce{C3H} &                 C1=C=[C]1 &                 13.4800 &    \citet{Gratier_2016} \\
     \ce{C3H} &              [CH+]=C=[C-] &                 12.7497 &    \citet{Gratier_2016} \\
    \ce{C3H2} &                   C1=C=C1 &                 11.7701 &    \citet{Gratier_2016} \\
    \ce{C3H2} &                   C1C\#C1 &                 13.2695 &    \citet{Gratier_2016} \\
    \ce{C3H2} &                   C=C=[C] &                 12.3979 & \citet{Cernicharo_1991} \\
  \ce{CH3CCH} &                     CC\#C &                 14.0607 &    \citet{Gratier_2016} \\
     \ce{C2O} &               [C+]\#C[O-] &                 12.5705 &    \citet{Gratier_2016} \\
  \ce{CH2CN} &                 [CH2]C\#N &                 13.5798 &    \citet{Gratier_2016} \\
  \ce{CH3CN} &                     CC\#N &                 12.6096 &    \citet{Gratier_2016} \\
    \ce{HNCO} &                     N=C=O &                 13.0294 &    \citet{Gratier_2016} \\
      \ce{CS} &                [C-]\#[S+] &                 13.4594 &    \citet{Gratier_2016} \\
  \ce{CH3CHO} &                      CC=O &                 12.4298 &    \citet{Gratier_2016} \\
    \ce{HCS+} &                   C\#[S+] &                 12.7597 &    \citet{Gratier_2016} \\
    \ce{H2CS} &                       C=S &                 13.6201 &    \citet{Gratier_2016} \\
      \ce{SO} &                       S=O &                 13.6702 &    \citet{Gratier_2016} \\
     \ce{C4H} &                [C]\#CC\#C &                 13.4298 &    \citet{Gratier_2016} \\
    \ce{C4H2} &                 C=C=C=[C] &                 13.3365 &    \citet{Gratier_2016} \\
     \ce{C3N} &                [C]\#CC\#N &                 13.5502 &    \citet{Gratier_2016} \\
    \ce{HNC3} &          [C-]\#C-C\#[NH+] &                 11.6803 &    \citet{Gratier_2016} \\
     \ce{C3O} &               [C]\#C[C]=O &                 11.9201 &    \citet{Gratier_2016} \\
  \ce{HC3NH+} &              C\#CC\#[NH+] &                 11.8698 &    \citet{Gratier_2016} \\
 \ce{CH2CHCN} &                   C=CC\#N &                 12.8102 &    \citet{Gratier_2016} \\
  \ce{HCCCHO} &                   C\#CC=O &                 11.2601 &    \citet{Gratier_2016} \\
     \ce{C2S} &               [C+]\#C[S-] &                 14.0086 &    \citet{Gratier_2016} \\
     \ce{OCS} &                     O=C=S &                 13.2601 &    \citet{Gratier_2016} \\
     \ce{C5H} &          [CH+]=C=C=C=[C-] &                 12.2695 &    \citet{Gratier_2016} \\
     \ce{C3S} &            [C-]\#CC\#[S+] &                 13.1399 &    \citet{Gratier_2016} \\
     \ce{C6H} &            [C]\#CC\#CC\#C &                 12.7404 &    \citet{Gratier_2016} \\
    \ce{HC3N} &                  C\#CC\#N &                 14.2430 &        \citet{Xue_2020} \\
  \ce{HCCNC} &            C\#C[N+]\#[C-] &                 12.5821 &        \citet{Xue_2020} \\
    \ce{HC5N} &              C\#CC\#CC\#N &                 13.8254 &        \citet{Xue_2020} \\
  \ce{HC4NC} &        C\#CC\#C[N+]\#[C-] &                 11.5172 &        \citet{Xue_2020} \\
    \ce{HC7N} &          C\#CC\#CC\#CC\#N &                 13.5623 &        \citet{Xue_2020} \\
  \ce{HC6NC} &    C\#CC\#CC\#C[N+]\#[C-] &                 11.6064 &        \citet{Xue_2020} \\
    \ce{HC9N} &      C\#CC\#CC\#CC\#CC\#N &                 13.3345 &     \citet{Loomis_2021} \\
  \ce{HC11N} &  C\#CC\#CC\#CC\#CC\#CC\#N &                 12.0170 &     \citet{Loomis_2021} \\
  \ce{C5H5CN} &             C1C=CC=C1C\#N &                 11.9191 & \citet{Kelvin_Lee_2021} \\
  \ce{C5H5CN} &           C1C=CC(=C1)C\#N &                 11.2788 & \citet{Kelvin_Lee_2021} \\
  \ce{C11H7N} & C1=CC=C2C(=C1)C=CC=C2C\#N &                 11.8663 &    \citet{mcguire_detection_2021} \\
  \ce{C11H7N} & C1=CC=C2C=C(C=CC2=C1)C\#N &                 11.8482 &    \citet{mcguire_detection_2021} \\
  \ce{C6H5CN} &         C1=CC=C(C=C1)C\#N &                 12.2380 &    \citet{mcguire_detection_2021} \\
\ce{HCCCH2CN} &                 C\#CCC\#N &                 11.9643 &    \citet{McGuire_2020} \\
  \ce{H3C5N} &             C\#C/C=C/C\#N &                 11.3874 & \citet{lee_discovery_2021} \\
  \ce{H3C5N} &               C=CC\#CC\#N &                 11.0719 & \citet{lee_discovery_2021} \\
  \ce{H3C5N} &             C\#C/C=C\textbackslash C\#N &                 11.3032 & \citet{lee_discovery_2021} \\
     \ce{C8H} &        [C]\#CC\#CC\#CC\#C &                 11.6628 &    \citet{Br_nken_2007} \\
    \ce{C8H-} &       C\#CC\#CC\#CC\#[C-] &                 10.3222 &    \citet{Br_nken_2007} \\
    \ce{C6H-} &           C\#CC\#CC\#[C-] &                 11.0792 &    \citet{Br_nken_2007} \\
    \ce{C4H-} &               C\#CC\#[C-] &                 10.9294 &    \citet{Br_nken_2007} \\
  \ce{H2CCO} &                     C=C=O &                 12.7118 &       \citet{Soma_2018} \\
      \ce{CN} &                    [C]\#N &                 12.8899 &     \citet{Pratap_1997} \\
     \ce{HNC} &               [C-]\#[NH+] &                 12.6201 &     \citet{Pratap_1997} \\
    \ce{HC7O} &        C\#CC\#CC\#C[C+]=O &                 11.8921 &   \citet{Cordiner_2017} \\
    \ce{HC5O} &            C\#CC\#C[C+]=O &                 12.2304 &    \citet{McGuire_2017} \\
    \ce{H2CN} &                     C=[N] &                 11.1761 &     \citet{Ohishi_1994} \\
    \ce{H2CO} &                       C=O &                 13.0792 &       \citet{Soma_2018} \\
  \ce{HC3O+} &               C\#CC\#[O+] &                 11.3222 & \citet{Cernicharo_2020} \\
  \ce{HOCO+} &                 O=C=[OH+] &                 11.6021 & \citet{Cernicharo_2020} \\
  \ce{H2COH+} &                   C=[OH+] &                 11.4771 & \citet{Cernicharo_2020} \\
  \ce{H2NCO+} &                  NC\#[O+] &                 10.6021 & \citet{Cernicharo_2020} \\
    \ce{HCNO} &                   C\#N[O] &                 10.8451 & \citet{Cernicharo_2020} \\
    \ce{HOCN} &                     OC\#N &                 11.0414 & \citet{Cernicharo_2020} \\
     \ce{C4O} &            [C]\#CC\#[C]=O &                 11.0792 & \citet{Cernicharo_2020} \\
  \ce{HCOOH} &                    C(=O)O &                 12.1461 & \citet{Cernicharo_2020} \\
    \ce{HC2O} &                  C\#[C]=O &                 12.0000 & \citet{Cernicharo_2020} \\
    \ce{HC3O} &                 C\#C[C]=O &                 11.3010 & \citet{Cernicharo_2020} \\
    \ce{HC4O} &              C\#CC\#[C]=O &                 11.4771 & \citet{Cernicharo_2020} \\
  \ce{H2C3O} &                   C=C=C=O &                 11.0414 & \citet{Cernicharo_2020} \\
  \ce{H2C3O} &               C1=C(=O)=C1 &                 11.6021 & \citet{Cernicharo_2020} \\
      \ce{CH} &                      [CH] &                 14.1461 &      \citet{Sakai_2013} \\
    \ce{CNCN} &                [C]\#NC\#N &                 11.9542 &    \citet{Ag_ndez_2018} \\
  \ce{NCCNH+} &              N\#CC\#[NH+] &                 10.9345 &    \citet{Ag_ndez_2015} \\
    \ce{C6H2} &             C=C=C=C=C=[C] &                 10.3284 &     \citet{Langer_1997} \\
\ce{CH3CHCH2} &                      CC=C &                 13.6021 &  \citet{Marcelino_2007} \\
\ce{CH2C2HCN} &                 C=C=CC\#N &                 11.6532 &      \citet{Lovas_2006} \\
     \ce{HCN} &                      C\#N &                 12.3892 &     \citet{Hirota_1998} \\
    \ce{C9H8} &          c1ccc2c(c1)CC=C2 &                 12.9823 &                     \cite{burkhardt_discovery_2021} \\
\ce{CH2CHCCH} &                   C=CC\#C &                 13.0792 & \citet{Cernicharo_2021} \\
    \ce{HCCN} &                 N\#C[CH+] &                 11.6435 & \citet{Cernicharo_2021} \\
\ce{CH3CH2CN} &                    CCC\#N &                 11.0414 & \citet{Cernicharo_2021} \\
\enddata
\end{deluxetable}

\end{document}